\documentclass[11pt]{article}
\usepackage{ifthen} \newboolean{isSpringer} \setboolean{isSpringer}{false}

\usepackage[utf8]{inputenc}

\usepackage{mathptmx}       
\usepackage{helvet}         
\usepackage{courier}        
\usepackage{type1cm}        

\usepackage{makeidx}         
\usepackage{graphicx}        
\usepackage{multicol}        
\usepackage[bottom]{footmisc}
\usepackage[normalem]{ulem}	
\usepackage{soul}   

\usepackage[utf8]{inputenc}
\usepackage{amsfonts,amsmath,amssymb} 
\usepackage{eurosym} 
\usepackage[defblank]{paralist} 
\usepackage[obeyspaces]{url} 
\usepackage[verbatim]{svn-multi}
\usepackage[dvipsnames]{xcolor}
\usepackage{xspace}

\usepackage[all]{xy}

\ifthenelse{\boolean{isSpringer}}{
}{
   \usepackage{leading} \leading{6mm}
   \usepackage[authoryear,round]{natbib}
   \setcounter{secnumdepth}{0}
   \usepackage{fancyhdr}
   \pagestyle{fancy} \fancyhf{} 
   \cfoot{\thepage{}}
   \bibpunct{(}{)}{;}{a}{}{,} 
}

\let\cite=\citep

\usepackage[
  draft,
  colorlinks = true,
  linkcolor = black,
  urlcolor  = black,
  citecolor = black,
  anchorcolor = black]{hyperref}

\def\land{\mathrel{\wedge}}

\begin{document}

\ifthenelse{\boolean{isSpringer}}{
  \title{Conflict-free Replicated Data Types}
  \title*{%
    Conflict-free Replicated Data Types%
  }
  \titlerunning{CRDTs}
  
  \author{Nuno Pregui{\c c}a
    \and Carlos Baquero
    \and Marc Shapiro}
  \authorrunning{Nuno Pregui{\c c}a et al.}
  \institute{Nuno Preguiça
    \at DI, FCT, Universidade NOVA de Lisboa and NOVA LINCS, Portugal. 
    \email{nuno.preguica@fct.unl.pt}
  \and Carlos Baquero
    \at HASLab / INESC TEC \& Universidade do Minho,
    Portugal.
    \email{cbm@uminho.pt}
  \and Marc Shapiro
    \at Sorbonne-Universités-UPMC-LIP6 \& Inria Paris,
    \email{http://lip6.fr/Marc.Shapiro/}}

}{
  \title{Conflict-free Replicated Data Types}
  \author{
    Nuno Pregui{\c c}a\\ {\small DI, FCT, Universidade NOVA de Lisboa and NOVA LINCS, Portugal}
    \and Carlos Baquero\\ {\small HASLab / INESC TEC \& Universidade do Minho, Portugal}
    \and Marc Shapiro\\ {\small Sorbonne-Université \& Inria, Paris, France}
  }
  \date{20 February 2018}
}
\maketitle

\section{Definition}
A conflict-free replicated data type (CRDT) is an abstract data type, with a well defined interface,
designed to be replicated at multiple processes and exhibiting the following properties:
\begin{inparaenum}[(i)]
\item any replica can be modified without coordinating with another replicas;
\item when any two replicas have received the same set of updates, they reach the same state, deterministically, 
by adopting mathematically sound rules to guarantee state convergence. 
\end{inparaenum}

\section{Overview}

Internet-scale distributed systems often replicate data at multiple geographic locations
to provide low latency and high availability, despite outages and network failures.
To this end, these systems must accept updates at any replica, and propagate these 
updates asynchronously to the other replicas. 
This approach allows replicas to temporarily diverge and requires a mechanism for merging 
concurrent updates into a common state. 
CRDTs provide a principled approach to address this problem.

As any abstract data type, a CRDT implements some given functionality and exposes a 
well defined interface. Applications interact with the CRDT only through this interface.
As CRDTs are specially designed to be replicated and to allow uncoordinated updates, a
key aspect of a CRDT is its semantics in the presence of concurrency. 
The concurrency semantics defines what is the behavior of the object in the presence of 
concurrent updates, defining the state of the object for any given set of received updates.

An application developer 
uses the CRDT interface 
and concurrency semantics to reason about the behavior of her applications in 
the presence of concurrent updates.
A system developer that needs to create a system that provides CRDTs needs to focus
on another aspect of CRDTs: the synchronization model. 
The synchronization model defines the requirements that the system must meet so that
CRDTs work correctly. We now detail each of these aspects independently.

\subsection{Concurrency semantics}

The operations defined in a data-type may intrinsically commute or not. 
Consider for instance a Counter data type, a shared integer that supports increment and decrement operations. 
As these operations commute (i.e., executing them in any order yields the same result) the Counter data 
type naturally converges towards the expected result.
In this case, it is natural that the state of a CRDT object reflects all 
executed operations.

Unfortunately, for most data-types, this is not the case and several 
concurrency semantics are reasonable, with different semantics being 
suitable for different applications.
For instance, consider a shared memory cell supporting the assignment operation. 
If the initial value is 0, the correct outcome for concurrently assigning 1 and 2 
is not well defined.

When defining the concurrency semantics, an important concept that is often used
is that of the \emph{happens-before} relation \citep{Lamport78Time}. 
In a distributed system, an event $e_1$ \emph{happened-before} an event $e_2$, 
$e_1 \prec e_2$, iff:
\begin{inparaenum}[(i)]
\item $e_1$ occurred before $e_2$ in the same process; or 
\item $e_1$ is the event of sending message $m$, and $e_2$ is the event
of receiving that message; or
\item there exists an event $e$ such that $e_1 \prec e$ and $e \prec e_2$. 
\end{inparaenum}
When applied to CRDTs, we can say that an
update $u_1$ \emph{happened-before} an update $u_2$ iff the effects of $u_1$ 
had been applied in the replica where $u_2$ was executed initially.

As an example, if an event was ``\emph{Alice reserved the meeting room}'' it is relevant to know if that was known when ``\emph{Bob reserved the meeting room}'' to determine if Alice should be given priority or if two users concurrently tried to reserve the same room. 

For instance, let us use \emph{happened-before} to define the semantics of the 
\emph{add-wins} set (also known as observed-remove set, OR-set \citep{Shapiro11Conflict}). 
Intuitively, in the \emph{add-wins} semantics, 
in the presence of two operations that do not commute, a concurrent add and remove of the same element, the 
add wins leading to a state where the element belongs to the set.
More formally, the set interface has two update operations: 
\begin{inparaenum}[(i)]
\item $\mathsf{add}(e)$, for adding element $e$ to the set; and 
\item $\mathsf{rmv}(e)$, for removing element $e$ from the set.
\end{inparaenum}
Given a set of update operations $O$ that are related by the happens 
before partial order $\prec$, the state of the set is defined as:
$\{e \mid \mathsf{add}(e) \in O \land \nexists \mathsf{rmv}(e) \in O \cdot \mathsf{add}(e) \prec \mathsf{rmv}(e)\}$.

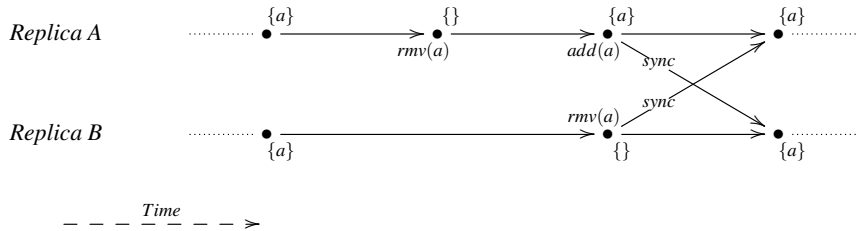
\begin{figure*}[h!]
\begin{center}
\footnotesize
\begin{xy}
\xymatrix{
  \mathit{Replica\; A} & \ar@{.}[r] &
  {\bullet} \ar@{->}[rr]^<{\{a\}}_>{rmv(a)} & & 
  {\bullet} \ar@{->}[rr]^<{\{\}}_>{add(a)} & &
  {\bullet} \ar@{->}[rr]^<{\{a\}}\ar@{->}[rrd] |(0.3){sync} & & 
  {\bullet} \ar@{.}[r]^<{\{a\}} &\\
  \mathit{Replica\; B} & \ar@{.}[r] &
  {\bullet} \ar@{->}[rrrr]_<{\{a\}}^>{rmv(a)} & & & & 
  {\bullet} \ar@{->}[rr]_<{\{\}} \ar@{->}[rru] |(0.3){sync} & & 
  {\bullet} \ar@{.}[r]_<{\{a\}} & \\
    \ar@{-->}[rr] ^{Time} & & &  & & & & &  
}
\end{xy} 
\end{center}
\caption{Run with an add-wins set.}
\label{fig:set:add-wins}
\end{figure*}

Figure~\ref{fig:set:add-wins} shows a run where an add-wins set is replicated
in two replicas, with initial state $\{a\}$. 
In this example, in replica A, $a$ is first removed and later added again to 
the set. In replica B, $a$ is removed from the set.
After receiving the updates from the other replica, both replicas end up 
with element $a$ in the set. The reason for this is that there is no
$rmv(a)$ that happened after the $add(a)$ executed in replica A.

An alternative semantics based on the happens-before relation is the \emph{remove-wins}.
Intuitively, in the \emph{remove-wins} semantics, 
in the presence of a concurrent add and remove of the same element, the 
remove wins leading to a state where the element is not in the set.
More formally, given a set of update operations $O$, the state of the set 
is defined as:
$\{e \mid \mathsf{add}(e) \in O \land \forall \mathsf{rmv}(e) \in O \cdot  \mathsf{rmv}(e) \prec \mathsf{add}(e)\}$.
In the previous example, after receiving the updates from the other replica, 
the state of both replicas 
would be the empty set, because there is no $add(a)$ that happened after the
$rmv(a)$ in replica B.

Another relation that can be useful for defining the concurrency semantics is 
that of a total order among updates and particularly a total order that 
approximates wall-clock time. 
In distributed systems, it is common to maintain nodes with their physical clocks 
loosely synchronized. 
When combining the clock time with a site identifier, we have unique 
timestamps that are totally ordered.
Due to the clock skew among multiple nodes, although these timestamps
approximate an ideal global physical time, they do not necessarily respect the happens-before
relation.
This can be achieved by combining physical and logical clocks, as
shown by Hybrid Logical Clocks \citep{DBLP:conf/opodis/KulkarniDMAL14}, or by only arbitrating a wall-clock total order for the events that are concurrent under causality \citep{Zawirski:2016:ECR:2911151.2911157}.

This relation allows to define the \emph{last-writer-wins} semantics, where the 
value written by the last writer wins over the values written previously, 
according to the defined total order.
More formally, with the set $O$ of operations now totally ordered by $<$, the state
of a \emph{last-writer-wins} set would be defined as:
$\{e \mid \mathsf{add}(e) \in O \land \forall \mathsf{rmv}(e) \in O \cdot \mathsf{rmv}(e) < \mathsf{add}(e)\}$.
Returning to our previous example, the state of the replicas after the
synchronization would include $a$ if, according the total order defined 
among the operations, the $rmv(a)$ of replica B is smaller than
the $add(a)$ of replica A. Otherwise, the state would be the empty set.

We now briefly introduce the concurrency semantics proposed for several 
CRDTs.

\subsubsection{Set}
For a set CRDT, we have shown the 
difference between three possible concurrency semantics: 
\emph{add-wins}, \emph{remove-wins} and \emph{last-writer-wins}. 

\subsubsection{Register}
A register CRDT maintains an opaque value and provides a single update
operation that writes an arbitrary value: $\mathsf{wr}(\mathit{value})$. 
Two concurrency semantics have been proposed leading to two different CRDTs:
the \emph{multi-value} register and the \emph{last-writer-wins} register.
In the \emph{multi-value} register, all concurrently written values are kept.
In this case, the read operation return the set of concurrently written values.
Formally, the state of a multi-value register is defined as the multi-set:
$\{v \mid \mathsf{wr}(v) \in O \land \nexists \mathsf{wr}(u) \in O \cdot \mathsf{wr}(v) \prec \mathsf{wr}(u)\}$.

In the \emph{last-writer-wins} register, only the value of the last write
is kept, if any. Formally, the state of a last-writer-wins register can be defined as a set that is either empty or holds a single value:
$\{v \mid \mathsf{wr}(v) \in O \land \nexists \mathsf{wr}(u) \in O \cdot \mathsf{wr}(v) < \mathsf{wr}(u)\}$.

\subsubsection{Counter}
A counter CRDT maintains an integer and can be modified 
by update operations $\mathsf{inc}$ and $\mathsf{dec}$, to increase and decrease by one unit its value, respectively
(this can easily generalize to arbitrary amounts).
As mentioned previously, as operations intrinsically commute, the natural
concurrency semantics is to have a final state that reflects the effects
of all registered operations. Thus the result state is obtained by counting the number of increments and subtracting the number of decrements: $\left| \{ \mathsf{inc} \mid \mathsf{inc} \in O \} \right| - \left| \{ \mathsf{dec} \mid \mathsf{dec} \in O \} \right|$. 

Now consider that we want to add a write operation $\mathsf{wr}(n)$, to update the value
of the counter to a given value. 
This opens two questions related with the concurrency semantics.
First, what should be the final state when two concurrent write operations
are executed. In this case, the last-writer-wins semantics would be simple 
(as maintaining multiple values, as in the multi-value register, is
overly complex).

Second, what is the result when concurrent writes and $\mathsf{inc}$/$\mathsf{dec}$ operations 
are executed. 
In this case, by building on the happens-before relation, we can 
define several concurrency semantics.  
One possibility is a \emph{write-wins} semantics, where $\mathsf{inc}$/$\mathsf{dec}$ 
operations have no effect when executed concurrently with the 
last write.
Formally, for a given set $O$ of updates that include at least a write operation, 
let $v$ be the value in the last write, i.e., $\mathsf{wr}(v) \in O \land \nexists \mathsf{wr}(u) \in O \cdot \mathsf{wr}(v) < \mathsf{wr}(u)$. The value of the counter would be
$v + o$, with $o = \left| \{ \mathsf{inc} \mid \mathsf{inc} \in O \wedge \mathsf{wr}(v) \prec \mathsf{inc}\} \right| - \left| \{ \mathsf{dec} \mid \mathsf{dec} \in O \wedge \mathsf{wr}(v) \prec \mathsf{dec}\} \right|$
representing $\mathsf{inc}$/$\mathsf{dec}$ operations that happened after the last write.

\subsubsection{Other CRDTs} 
A number of other CRDTs have been proposed in literature, including 
CRDTs for elementary data structures, such as Lists \citep{Preguica09Commutative,Weiss09Logoot,Roh11Replicated}, 
Maps \citep{Brown14Riak,DBLP:journals/jpdc/AlmeidaSB18} and Graphs \citep{Shapiro11Conflict}, and more
complex structures, such as JSON documents \citep{Kleppmann17JSON}.
For each of these CRDTs, the developers have defined and implemented a type-specific 
concurrency semantics.

\subsection{Synchronization Model}

A replicated system needs to synchronize its replicas, by propagating and applying
updates in every replica. 
There are two main approaches to propagate updates: state-based 
and operation-based replication.

In state-based replication, replicas synchronize by establishing bi-directional (or
unidirectional) synchronization sessions, where both (one, resp.) replicas send
their state to a peer replica. 
When a replica receives the state of a peer, it merges the received state with 
its local state.
As long as the synchronization graph is connected, every update will eventually
propagate to all replicas.  

CRDTs designed for state-based replication define a merge function to 
integrate the state of a remote replica.
It has been shown \citep{Shapiro11Conflict} that all replicas of a CRDT converge if: 
\begin{inparaenum}[(i)] 
\item the states of the CRDT are partially ordered according to $\leq$ forming a 
join semilattice;
\item an operation modifies the state $s$ of a replica by an inflation, producing a new state that is
larger or equal to the original state according to $\leq$, 
  i.e., for any operation $m$, $s \leq m(s)$;
\item the merge function produces the join (least upper bound) of two states, 
i.e. for states $s,u$ it derives $s \sqcup u$. 
\end{inparaenum}

In operation-based replication, replicas converge by propagating operations 
to every other replica.
When an operation is received in a replica, it is applied to the local replica
state.
Besides requiring that every operation is reliably delivered to all replicas,
e.g. by using some reliable multicast communication subsystem, some systems
may require operations to be delivered according to some specific order,
with causal order being the most common.

CRDTs designed for operation-based replication must define, for each operation, 
a generator and an effector function. 
The generator function executes in the replica 
where the operation is submitted, it has no side-effects and generates a effector 
that encodes the side-effects of the operation. In other words, the effector is a closure created by the generator depending on the state of the origin replica.
The effector operation must be reliably executed in all replicas, where it
updates the replica state. 
\citet{Shapiro11Conflict} show that if effector operations are delivered
in causal order, replicas will converge to the same state if concurrent effector operations
commute.
If effector operations may be delivered without respecting causal order, then all 
effector operations must commute.
Most operation-based CRDT design require causal delivery. 

\emph{Alternative models:}
When operations modify only part of the state, propagating the complete 
state for synchronization to a remote replica is inefficient, as the remote 
replica already knows most of the state. 
Delta-state CRDTs \citep{DBLP:journals/jpdc/AlmeidaSB18} address this issue 
by propagating only delta-mutators, that encode the changes that have been made
to a replica since the last communication.
The first time a replica communicates with some other replica, the full state
needs to be propagated.
This can be further improved, as shown in big delta state 
CRDTs \citep{vanderLinde06bigdelta}, typically at the cost of storing more metadata
in the CRDT state. Another improvement is to compute digests that help determine which parts of a remote state are needed, avoiding shipping full states \citep{Enes2017}.  

In the context of operation-based replication, effector operations should be applied immediately in the source replica, that executed the generator. However, propagation to other replicas can be deferred for some period and effectors stored in an outbound log, presenting an opportunity to 
compress the log by rewriting some operations
-- e.g. two $add(1)$ operations
in a counter can be converted in a $add(2)$ operation. 
This mechanism has been used by Cabrita et. al. \citep{Cabrita17Non}. 
Delta-mutators can also be seen as a compressed representation
of a log of operations.

The operation-based CRDTs require executing a generator function against
the replica state to compute an effector operation.
In some scenarios, this may introduce an unacceptable delay for propagating
an operation. 
Pure-operation based CRDTs \citep{Baquero14Making} address this issue by 
allowing the original operations to be propagated to all replicas,
typically at the cost of more complex operations and of having to store 
more metadata in the CRDT state.

\section{Key research findings}

\subsection{Preservation of sequential semantics}
When modeling an abstract data type that has an established semantics under sequential execution, CRDTs should preserve that semantics. For instance, CRDT sets should ensure that if the last operation in a sequence of operations to a set added a given element, then a query operation immediately after that one will show the element to be present on the set. Conversely, if the last operation removed an element, then a subsequent query should not show its presence. 

Sequential execution can occur even in distributed settings if synchronization is frequent. An instance can be updated in replica A, merged into another replica B and updated there, and merged back into replica A before A tries to update it again. In this case we have a sequential execution, even though updates have been executed in different replicas.

Historically, not all CRDT designs have met this property. The \emph{two-phase set} CRDT (2PSet), does not allow re-adding an element that was removed, and thus it breaks the common sequential semantics. Later CRDT set designs, such as \emph{add-wins} and \emph{remove-wins} sets, do preserve the original sequential semantics while providing different arbitration orders to concurrent operations.

\subsection{Extended behaviour under concurrency}
Some CRDT designs handle concurrent operations by arbitrating a given sequential ordering to accommodate concurrent execution. For example, the state of a \emph{last-writer-wins} set replica shown by its interface can be explained by a a sequential execution of the operations according to the LWW total order used. When operations commute, such as in G-Counters and PN-Counters, there might even be several sequential executions that explain a given state. 

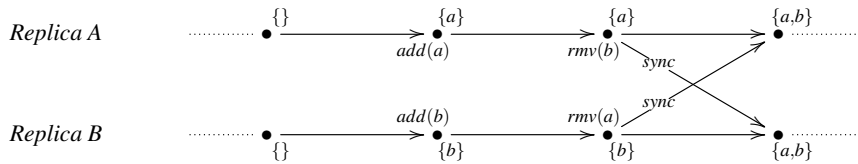
\begin{figure*}[h!]
\begin{center}
\footnotesize
\begin{xy}
\xymatrix{
  \mathit{Replica\; A} & \ar@{.}[r] &
  {\bullet} \ar@{->}[rr]^<{\{\}}_>{add(a)} & & 
  {\bullet} \ar@{->}[rr]^<{\{a\}}_>{rmv(b)} & &
  {\bullet} \ar@{->}[rr]^<{\{a\}}\ar@{->}[rrd] |(0.3){sync} & & 
  {\bullet} \ar@{.}[r]^<{\{a,b\}} &\\
  \mathit{Replica\; B} & \ar@{.}[r] &
  {\bullet} \ar@{->}[rr]_<{\{\}}^>{add(b)} & &
  {\bullet} \ar@{->}[rr]_<{\{b\}}^>{rmv(a)} & &
  {\bullet} \ar@{->}[rr]_<{\{b\}} \ar@{->}[rru] |(0.3){sync} & & 
  {\bullet} \ar@{.}[r]_<{\{a,b\}} &
}
\end{xy} 
\end{center}
\caption{Add-wins set run showing that there might be no sequential execution of operations that explains CRDTs behavior.}
\label{fig:set:add-wins:noseq}
\end{figure*}

Not all CRDTs need or can be explained by sequential executions. 
The add-wins set is an example of a CRDT where there might be no sequential execution of operations to explain the state observed, as Figure~\ref{fig:set:add-wins:noseq} shows.
In this example, the state of the set after all updates propagate to all replicas includes $a$ and $b$, but in any sequential extension of the causal order a remove operation would always be the last operation, and consequently the removed element could not belong to the set.

Some other CRDTs can exhibit states that are only attained when concurrency does occur. An example is the \emph{multi-value register}, a register that supports a simple write and read interface. If used sequentially, sequential semantics is preserved, and a read will show the outcome of the most recent write in the sequence. However if two or more value are written concurrently, the subsequent read will show all those values (as the \emph{multi-value} name implies), and there is no sequential execution that can explain this result. We also note that a follow up write can overwrite both a single value and multiple values.

\subsection{Guaranties and limitations}

An important property of CRDTs is that an operation can always be accepted at any given replica
and updates are propagated asynchronously to other replicas.
In the CAP theorem framework \citep{syn:rep:1648,rep:pan:1628}, the CRDT conflict-free approach 
favors availability over consistency when facing communication disruptions. 
This leads to resilience to network failure and disconnection, since no prior coordination
with other replicas is necessary before accepting an operation.
Further, operations can be accepted with minimal user perceived latency since 
they only require local durability.
By eschewing global coordination, replicas evolve independently and reads will 
not reflect operations accepted in remote replicas that have not yet been propagated 
to the local replica.


In the absence of global coordination, session guaranties \citep{DBLP:conf/pdis/TerryDPSTW94} 
specify what the user applications can expect from their interaction with the system's interface. 
Both state based CRDTs, and operation based CRDTs when supported by reliable causal delivery, 
provide per-object causal consistency. 
Thus, in the context of a given replicated object, the traditional session guaranties are met. 
CRDT based systems that lack transactional support can enforce system-wide causal consistency, 
by integrating multiple objects in a single map/directory object \citep{DBLP:journals/jpdc/AlmeidaSB18}. 
Another alternative is to use mergeable transactions to read from a causally-consistent database snapshot
and to provide write atomicity \citep{DBLP:conf/srds/PreguicaZBDBBS14}. 

Some operations cannot be expressed in a conflict free framework and will require global agreement. As an example, in an auction system, bids can be collected under causal consistency, and a new bid will only have to increase the offer with respect to bids that are known to causally precede it. However, closing the auction and selecting a single winning bid will require global agreement. It is possible to design a system that integrates operations with different coordination requirements and only resorts to global agreement when necessary \citep{DBLP:conf/osdi/LiPCGPR12,Sovran11Transactional}. 

Some global invariants, that usually are enforced with global coordination, can be enforced in 
a conflict free manner by using escrow techniques \cite{ONeil86Escrow} that split the available resources
by the different replicas.  
For instance, the Bounded Counter CRDT \citep{DBLP:conf/srds/BalegasSDFSRP15} defines a counter that never goes 
negative, by assigning to each replica a number of allowed decrements under the condition that 
the sum of all allowed decrements do not exceed the value of the counter.
While its assigned decrements are not exhausted, replicas can accept decrements without coordinating
with other replicas.
After a replica exhaust its allowed decrements, a new decrement will either fail or require synchronizing 
with some replica that still can decrement.
This technique uses point to point coordination, and can be generalized to enforce other system wide invariants \citep{DBLP:conf/eurosys/BalegasDFRPNS15}.

\section{Examples of applications}

CRDTs have been used in a large number of distributed systems and applications
that adopt weak consistency models. 
The adoption of CRDTs simplifies the development of these systems and applications, 
as CRDTs guarantee that replicas converge to the same state when all updates are 
propagated to all replicas.
We can group the systems and applications that use CRDTs into two groups: storage
systems that provide CRDTs as their data model; and applications that embed CRDTs
to maintain their internal data.

CRDTs have been integrated in several storage systems that make them available 
to applications.
An application uses these CRDTs to store their data, being the responsibility of 
the storage systems to synchronize the multiple replicas.
The following commercial systems use CRDT: Riak,\footnote{Developing with Riak KV 
Data Types \url{http://docs.basho.com/riak/kv/2.2.3/developing/data-types/}.} Redis \citep{rediscrdts} and Akka.\footnote{Akka Distributed Data: \url{https://doc.akka.io/docs/akka/2.5.4/scala/distributed-data.html}.}
A number of research prototypes have also used CRDT, including Walter~\citep{Sovran11Transactional},
SwiftCloud~\citep{DBLP:conf/srds/PreguicaZBDBBS14} and Antidote\footnote{Antidote:\url{http://antidotegb.org/}.}~\citep{Akkoorath16Cure}.

CRDTs have also been embedded in multiple applications. 
In this case, developers either used one of the available CRDT libraries,
implemented themselves some previously proposed design or designed new
CRDTs to meet their specific requirements. 
An example of this latter use is Roshi\footnote{Roshi is a large-scale CRDT set implementation for timestamped events \url{https://github.com/soundcloud/roshi}.}, a LWW-element-set CRDT used 
for maintaining an index in SoundCloud stream.

\section{Future directions of research}

\subsection{Scalability}
In order to track concurrency and causal predecessors, CRDT implementations often store metadata that grows linearly with the number of replicas \citep{DBLP:journals/ipl/Charron-Bost91}. While global agreement suffers from greater scalability limitations since replicas must coordinate to accept each operation, the metadata cost from causality tracking can limit the scalability of CRDTs when aiming for more than a few hundred replicas. A large metadata footprint can also impact on the computation time of local operations, and will certainly impact the required storage and communication. 

Possible solutions can be sought in more compact causality representations when multiple replicas are synchronized among the same nodes \citep{DBLP:journals/dc/MalkhiT07,DBLP:conf/srds/PreguicaZBDBBS14,DBLP:conf/srds/GoncalvesABF17} or by hierarchical approaches that restrict all to all synchronization and enable more compact mechanisms \citep{DBLP:journals/corr/AlmeidaB13}. 

\subsection{Reversible computation}
Non trivial Internet services require the composition of multiple sub-systems, to provide storage, data dissemination, event notification, monitoring and other needed components. When composing sub-systems, that can fail independently or simply reject some operations, it is useful to provide a CRDT interface that undoes previously accepted operations. 
Another scenario that would benefit from undo is collaborative editing of shared documents, where undo is typically a feature available to users.

Undoing an increment on a counter CRDT can be achieved by a decrement. Logoot-Undo \citep{Weiss10LogootUndo} proposes a solution for undoing (and redoing) operations for a sequence CRDT used for collaborative editing. However providing an uniform approach to undoing, reversing, operations over the whole CRDT catalog is still an open research direction. The support of undo is also likely to limit the level of compression that can be applied to CRDT metadata. 

\subsection{Security}
While access to a CRDT based interface can be restricted by adding authentication, any accessing replica has the potential to issue operations that can interfere with the other replicas. For instance, delete operations can remove all existing state. In state based CRDTs, replicas have access to state that holds a compressed representation of past operations and metadata. By manipulation of this state and synchronizing to other replicas, it is possible to introduce significant attacks to the system operation and even its future evolution. 

Applications that store state on third party entities, such as in cloud storage providers, might elect not to trust the provider and choose end-to-ends encryption of the exchanged state. This, however, would require all processing to be done at the edge, under the application control. A research direction would be to allow some limited form of computation, such as merging state, over information whose content is subject to encryption. Potential techniques, such as homomorphic encryption, are likely to pose significant computational costs. 
An alternative is to execute operations in encrypted data without disclosing it, relying on specific
hardware support, such as Intel SGX and ARM TrustZone.

\subsection{Non-uniform replicas}
The replication of CRDTs typically assumes that eventually all replicas
will reach the same state, storing exactly the same data.
However, depending on the read operations available in the CRDT interface, 
it might not be necessary to maintain the same state in all replicas.
For example, an object that has a single read operation returning the 
top-K elements added to the object only needs to maintain those top-K elements 
in every replica. 
The remaining elements are necessary if a remove operation is available,
as one of the elements not in the top needs to be promoted when a top element 
is removed. 
Thus, each replica can maintain only the top-K elements and the elements
added locally.

This replication model is named non-uniform replication \citep{Cabrita17Non}
and can be used to design CRDTs that exhibit important storage and bandwidth 
savings when compared with alternatives that keep all data in all replicas.
Although it is clear that this model cannot be used for all data types, 
several useful CRDT design have been proposed, including top-K, top-Sum and
histogram. To understand what data types can adopt this model and how to 
explore it in practice is an open research question.

\subsection{Verification}

An important aspect related with the development of distributed systems that 
use CRDTs is the verification of the correctness of the system. 
This involves not only verifying the correctness of CRDT designs, but also 
the correctness of the system that uses CRDTs. A number of works have addressed 
these issues. 

Regarding the verification of the correctness of CRDTs, several approaches have been 
taken. The most commonly used approach is to have proofs when designs are proposed
or to use some verification tools for the specific data type, such as 
TLA \citep{Lamport94TLA} or Isabelle\footnote{Isabelle: \url{http://isabelle.in.tum.de/}.}.
There has also been some works that proposed general techniques 
for the verification of CRDTs \citep{Burckhardt14Replicated,Zeller14Formal,Gomes17Verifying}, 
which can be used by CRDT developers to verify the correctness of their designs.
Some of these works \citep{Zeller14Formal,Gomes17Verifying} 
include specific frameworks that help the developer 
in the verification process.

A number of other works have proposed techniques to verify the correctness
of distributed systems that use CRDTs \citep{Gotsman16Cause,Zeller17Testing,DBLP:conf/eurosys/BalegasDFRPNS15}.
These works typically require the developer to specify the properties that the
distributed system must maintain, and a specification of the operations in the 
system (that is independent of the actual code of the system). 
Despite these works, the verification of the correctness of CRDT designs and
of systems that use CRDTs, how these verification techniques can be made available
to programmers, and how to verify the correctness of implementations, remain 
an open research problem.

\section*{Acknowledgments}

This work was partially supported
by NOVA LINCS (UID/CEC/04516/2013), EU H2020 LightKone project
(732505), and SMILES line in project TEC4Growth (NORTE-01-0145-FEDER-000020).

\bibliographystyle{spbasic}
\bibliography{crdts}

\end{document}